\documentclass[10pt,a4paper,oneside,twocolumn]{article}

\usepackage{style/MyAdd}

\author{
\and
\\
$^1$ INFN-Torino\\ 
$^2$ Thales Alenia Space-Italy\\ 
$^3$ INFN-Trieste\\
\\
falzetta@to.infn.it 
}

\title{GEANT4 and CREME96 comparison using only proton fluxes}

\begin{document}

\maketitle

\abstract{
{\em
CREME96 and GEANT4 are two well known particle transport codes through matter in space science.
We present a comparison between the proton fluxes outgoing from an aluminium target, obtained by using both tools. The primary proton flux is obtained by CREME96 only, covering an energy range from MeV to hundreds GeV with the same result in both cases.
We studied different thickness targets and two different GEANT4 physics lists in order to show how the spectra of the outgoing proton fluxes are modified. 
Our findings show consistent agreement of simulation data by each tool, with regards both GEANT4 physics lists and every thickness target analysed.
}

}

\section*{Introduction}
The scientific world makes use of several simulation codes to study the interactions between radiation and matter, as can be see in the overview reported in sec.~\ref{StateOfArt}.
In our study, we focused on two of these, CREME96 and GEANT4 both described in sec.~\ref{Simulators}.
These two toolkits are characterized by different physics, both illustrated in sec.~\ref{physic:Simulators}.
In order to realize this study we implemented a specific simulation set-up, as defined in sec.~\ref{Simulations}, and we used a statistic  data analysis explained in sec.~\ref{SimulationsAnalysis}.
Our results, reported in sec.~\ref{Results}, show the agreement between CREME96 and GEANT4. 
As we suggest in \ref{FeaturesWorks}, these promising results have encouraged us to undertake further comparisons using different types of primary particles.

\section{State of Art}
Nowadays there are several particle transport codes through matter, for example CREME86, CREME96\cite{CREME96}, FLUKA\cite{FLUKA}, GEANT3\cite{GEANT3}, GEANT4\cite{GEANT4}, MARS\cite{MARS}, MCNP\cite{MCNP}, MCNPX\cite{MCNPX}, PHITS.

These tools use different languages, but most of the codes are designed in fortran; f77 for GEANT3, FLUKA, PHITS; f90 in the case of MCNPX and f95, for MARS. 
The only one of these examples in C++ is GEANT4. 
Source codes for CREME86 and 96 are not freely available.

Each is developed by a different community. 
The owner of CREME, 86 and 96, is the USA Naval Research Laboratory. 
On the other hand MCNPX is developed by LANL. 
The GEANT4 community includes CERN, ESA, IN2P3, PPARC, INFN, LIP, KEK, SLAC, TRIUMF. 
FLUKA support is made by CERN and INFN researchers and MARS is a FNAL laboratory production.
Finally, PHITS code is managed by JAEA, RIST, GSI and Chalmers University. 

Because of these differences, working simultaneously with all these codes is a very hard task. 
Therefore a scientist usually carries out research on one particular code and sometimes makes a comparison between their own software and another code. 
Examples of this kind of comparative study~\cite{BWR06} have already beeen undertaken for FLUKA and GEANT4 and~\cite{ID86} for CREME~86 and GEANT4.

\label{StateOfArt}

\section{Simulation tools considered}
As previously stated (sec.~\ref{StateOfArt}), there are many tools simulating the passage of particles through matter.
In our study we focused on two of these: CREME96 and GEANT4.

\label{Simulators}

    \begin{itemize}
    %\subsection{CREME96}
    \item Cosmic Ray Effects on Micro-Electronics (CREME) is a toolkit developed to study the effects of an ionizing radiation environment on the electronics inside a satellite, i.e. Single Event Upsets~(SEU) due to solar heavy ions~\cite{TD+96}.

\noindent The CREME96 user manages the simulation by {\em point and click} method, by means of an easy-to-use graphical interface avaiable on-line at~\cite{CREME96}.

\noindent CREME96's principal features, for instance the physical models used and their limitations, are reported in~\cite{TA+97}.
These models and their constraints are an important element of our study, for this reason we will describe them in more detail below.

\label{CREME96}

    %\subsection{Geant4}
    \item GEANT4, GEometry ANd Tracking (particles), is a Monte Carlo toolkit for simulating the passage of the particles through matter, avaiable on-line at~\cite{GEANT4}.

\noindent GEANT4's first production release was completed in December 1998~\cite{Gea03}. Unlike its ancestor GEANT3, in fortran~77,  GEANT4 is designed in C++.

\noindent In GEANT4 there is not a default physics list, therefore the user must define his/her specific physics.
The features of our GEANT4 physics list are explained in sec.~\ref{physic:geant4}.

\noindent Moreover, in GEANT4 there is no pre-implemented definition of particle flux. 
In order to overcome such a problem, QinetiQ developed MUlti-LAyered Shielding SImulation Software (MULASSIS~\cite{mulassis}), a GEANT4 interfacing tool, which also contains other features to work with this Monte Carlo code.

\noindent We did not consider MULASSIS for two reasons. 
Firstly, because we prefered to make a direct comparison between CREME96 and GEANT4 without being compelled to also take into account the additional aspect of an intermediate programme.
Secondly, we needed to input both CREME96 and GEANT4 radiation transport calculation packages with exactly the same particle flux output (the same particle flux output produced as a text file by CREME96).
Conducting the same study using MULASSIS would have added extra complications. 

\noindent For these two reasons we did not use MULASSIS, deciding instead to work with a simple code suitable for our research purposes.

\label{Geant4}
    \end{itemize}

\section{Physics in the two simulators}
CREME96 and GEANT4 were developed by two different scientific communities in different time periods. As a result their physics show a number of differences.
\label{physic:Simulators}

    %\subsection{The Physic in CREME96}
    \begin{itemize}

\item The physics of CREME96 includes models of the Galactic Cosmic Rays~(GCRs), Anomalous Cosmic Ray~(ACRs), Solar Particle Events~(SPEs) and a geomagnetic transmission calculations. These models are used for building a flux of particles about an orbit defined by the user. The proton flux used in our study is generated by CREME96 only, because it is not possible for GEANT4 to do it.

\noindent Another aspect of CREME96's physics model is the nuclear transport physics through the shielding material. 
This tool takes into account both ionization energy loss (in the continuous slowing down approximation) and nuclear fragmentation.
It implements stopping power and range-energy routines and uses semi-empirical energy-dependent nuclear fragmentation cross-sections\cite{TA+97}.

\noindent The physics in CREME96 has three limitations.
First of all, it does not consider neutron production in the shield, because these particles do not cause damage to electronics of spacescraft. 
Secondly, it does not include target fragments. 
Finally, aluminium is the only type of shielding material.

\item CREME96 can describe both the external environment (no shield) and the internal satellite (shielded) via flux of particles with Z range from 1 to 28, without requiring its user to define or to implement their flux definition.

\end{itemize}
\label{physic:creme96}

    %\subsection{The Physic in GEANT4}
    \begin{itemize}

%\subsubsection*{Phisic List}
\item With regards GEANT4, in this study we used two types of physic list, named {\em type A} and {\em type B}, in order to see the differences between the two simulation codes, but also between our two particular choices.

\noindent As far as the electromagnetic physics list was concerned both types included the following models: photon-epdl; electron-eedl; positron-standard; ion-ICRU; muon-standard and decay.

\noindent The difference between the two types is in the hadronic physics list selected, Binary for {\em type~A} and Bertini for {\em type~B}.

\noindent The physics implementation used in our study depended on many elements; in the case of GEANT4, for example, its patch and its library version.
In our particular study we used Geant4 version 8.1 patch 1 and CLHEP library version 1.9.2.3..

%\subsubsection*{Flusso}
\item As previously stated, GEANT4 does not have a default definition of particle flux, therefore we had to solved it in the first instance.
In our implementation of the particle flux ($\Phi$) in order to make a comparison for CREME96 $\Phi$ was defined as follows~\cite{ECSS}:

\begin{Def}[Flux]\label{def:Flux}
``amount of radiation crossing a surface per unit of time, often expressed in ``integral form'' as particles per unit area per unit time (e.g. electrons $cm^{-2}s^{-1}$) above a certain threshold energy.

NOTE The directional flux is the differential with respect to solid angle (e.g. particles $cm^{-2} \ steradian^{-1} \ s^{-1}$) while the ``differential'' flux is differential with respect to energy (e.g. particles $cm^{-2} \ MeV^{-1} \ s^{-1}$). In some cases fluxes are also treated as a differential with respect to Linear Energy Transfer (\dots).''
\end{Def}

In this paper, according to def.\ref{def:Flux}, the measurement unit of the flux used in the input file is:

\begin{center}
\begin{eqnarray}\label{def:Corrente}
\frac{particles}{m^2 \cdot s \cdot sr \cdot MeV/nucl}
\end{eqnarray}
\end{center}

where {\em m} is meter, {\em s} is second and {\em sr} is steradian and it
is normalized by the number of $primary$ particles generated during the simulation.

\end{itemize}\label{physic:geant4}

\section{Simulation set-up}
The considered geometric configuration and the material set up of the experiment were reproduced in both simulators in the same way and consisted of two elements: the beam and the target.

\begin{figure}[t]
  \centering
  \includegraphics[angle=0, width=12cm]{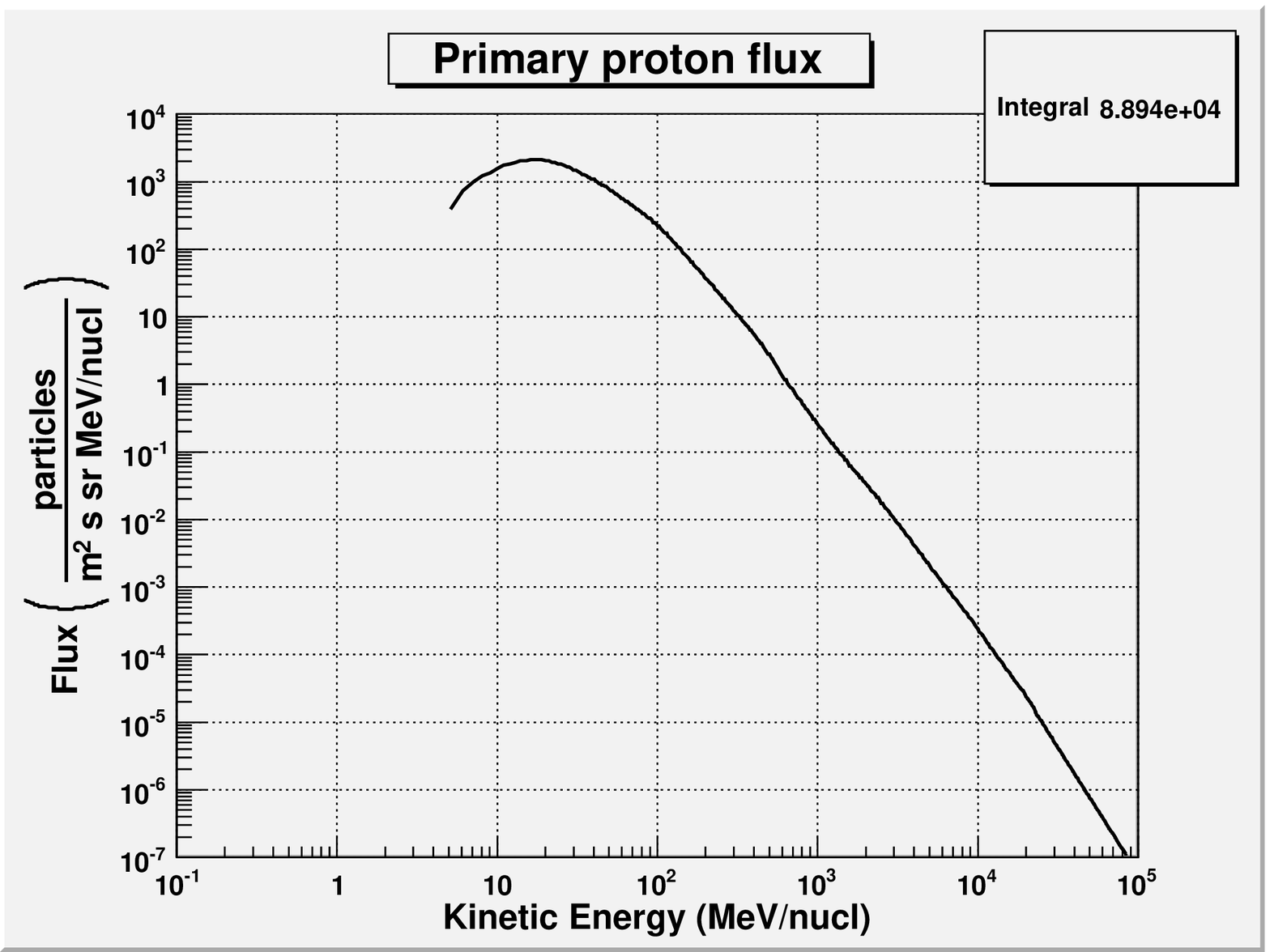}
  \caption{Primary proton fluxes,
           plotted in logarithmic scale.}\label{fig:primary}
\end{figure}

\begin{itemize}

\item The Beam. A punctiform beam made of protons was aimed at the face of the target, towards its center.

\noindent As previously stated (ref. sec.~\ref{physic:geant4}) the number of primary particles generated during the simulation can not be explained in this study, because the flux~$\Phi$ (ref. def.~\ref{def:Flux}) was normalized by primary particle number.

\noindent The energy weight of a single primary proton depends on the flux used with input for both simulations.
This flux is generated only by CREME96, as explained in sec.~\ref{physic:creme96}.
For the purpose of our paper, it derived from the choice of a typical ISS orbit thus defining its component environmental features (see fig.~\ref{fig:primary}).

\item The target. This was made of aluminium because, as previously stated (ref. sec.~\ref{physic:creme96}), CREME96 can only functions with this material.

\noindent The target thickness is made to vary along the line normal to the beam direction, from 1-5~cms at 1~cm intervals, in order to observe how the outgoing proton flux was modified.
These thickness values were chosen because they are typically used when modelling the satellite shielding action in a simple way, e.g.: FOTON-M 3~\cite{FOTON} can be approximated with a 3~cm layer of equivalent aluminium.

\end{itemize}

\label{Simulations}

\section{Simulation Analysis}
We limited our study to the proton fluxes both for simulator input and simulator output. In particular, we clarified that in the case of output in the particle flux there were both primary protons going out of the target and secondary protons produced in the shield.

\noindent CREME96's fluxes were not normalized by the number of primary particles, therefore in order to make a comparison we had to first perform normalization. 
For this tool we calculated the primary number by integration of the proton flux used for simulation input. 
The result was $8.894 \cdot 10^4$ particles, and is reported in fig.~\ref{fig:primary}.

\noindent The data output of the two simulators did not have the same statistics.

\noindent CREME96 data did not have the graphical error bar, whereas GEANT4 simulations received the following statistic:

\label{SimulationsAnalysis}

    %\subsection{The Statistic in GEANT4}
    x-axis error bars were constant and are equal to the bin\-ning chosen by the user during the implementation code.
In this study we have defined a 1 MeV/Nucleon bin.

The y-axis error bars were variable. 
The dimension of this bar type followed the Poisson statistic i.e.:

\begin{center}
\begin{eqnarray}\label{fig:ErrorBar}
\frac{1}{\sqrt{N}}
\end{eqnarray}
\end{center}

where $N$ is the number of entries w.r.t. the relative energetic bin.

\label{Statistic:geant4}

\section{Results and comparison}
\begin{figure}[t]
  \centering
  \includegraphics[angle=0, width=12cm]{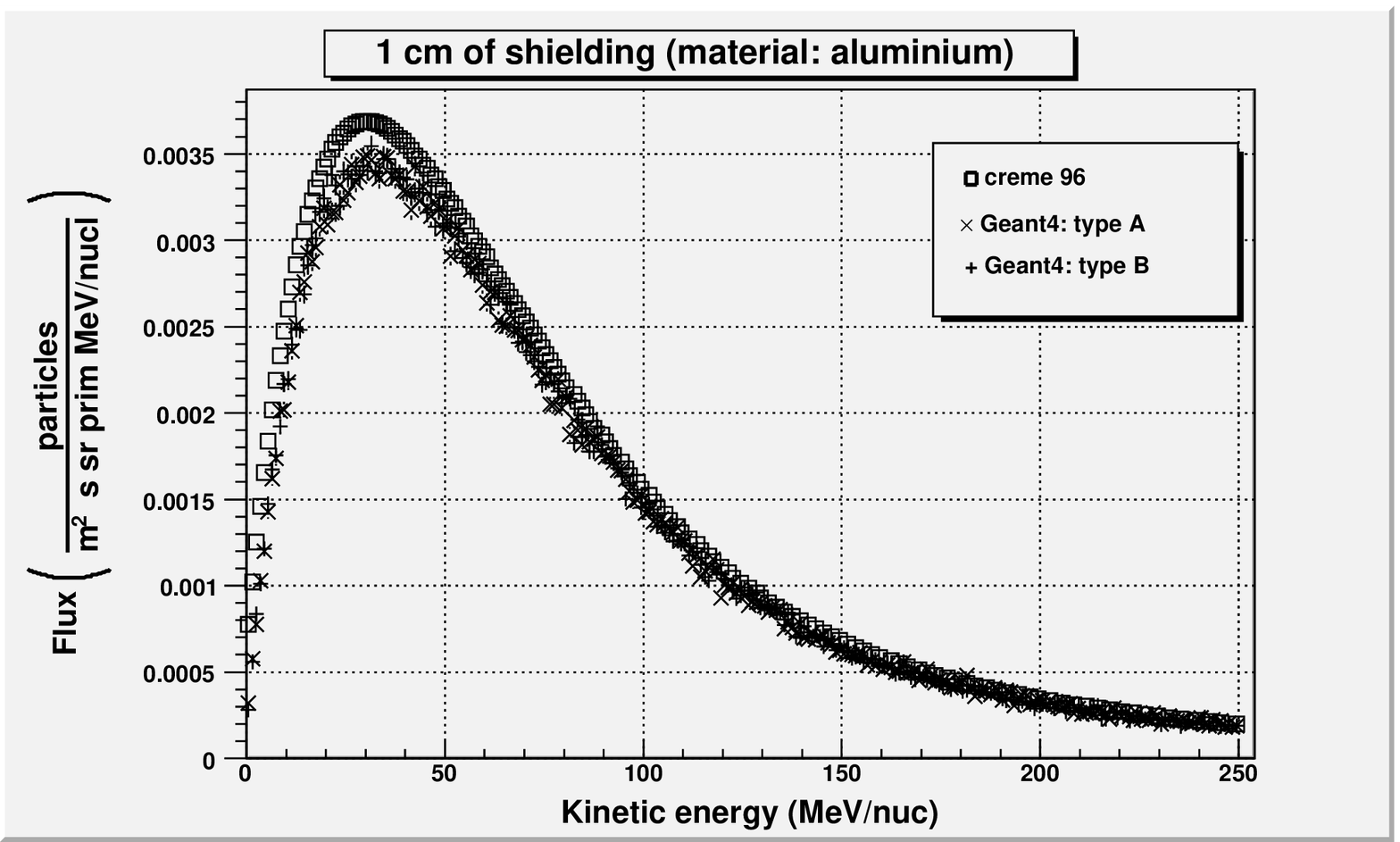}
  \caption{Proton fluxes behind 1~cm of aluminium shielding.
	   plotted in decimal scale.}\label{fig:1cm_dec}
\end{figure}

\begin{figure}[t]
  \centering
  \includegraphics[angle=0, width=12cm]{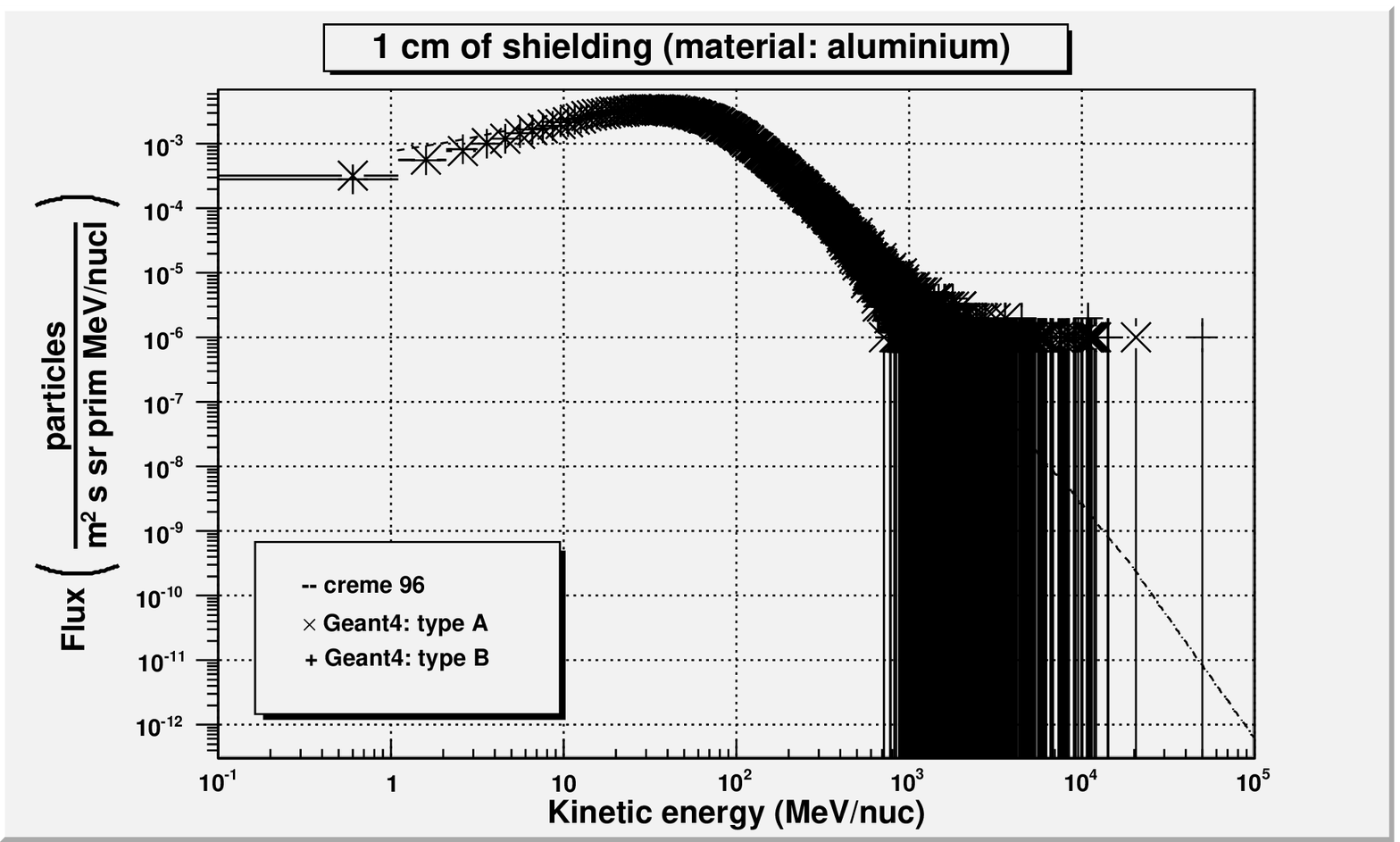}
  \caption{Proton fluxes behind 1~cm of aluminium shielding.
	   plotted in logarithmic scale.}\label{fig:1cm_log}
\end{figure}

\begin{figure}[t]
  \centering
  \includegraphics[angle=0, width=12cm]{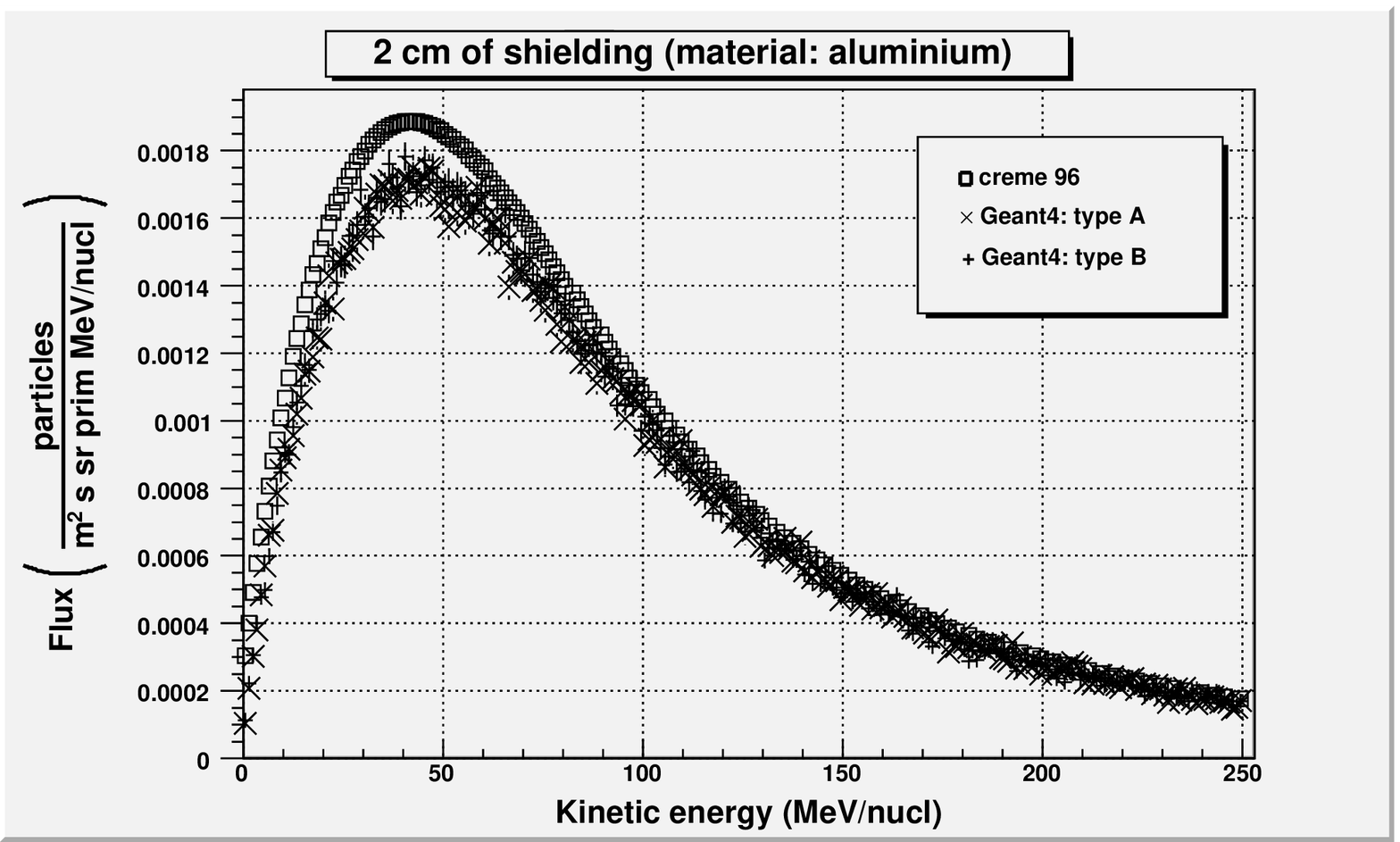}
  \caption{Proton fluxes behind 2~cm of aluminium shielding.
	   plotted in decimal scale.}\label{fig:2cm_dec}
\end{figure}

\begin{figure}[t]
  \centering
  \includegraphics[angle=0, width=12cm]{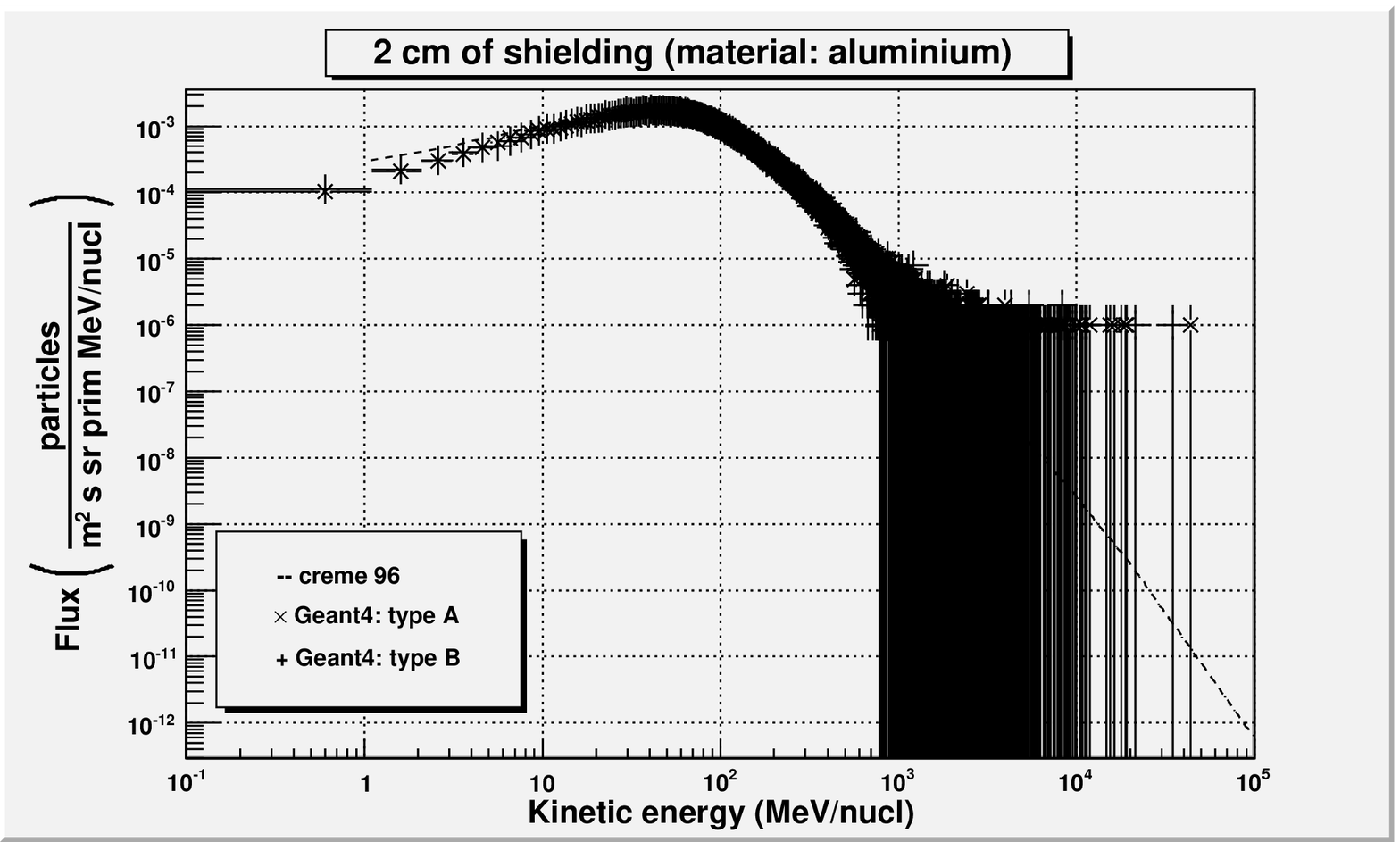}
  \caption{Proton fluxes behind 2~cm of aluminium shielding.
	   plotted in logarithmic scale.}\label{fig:2cm_log}
\end{figure}

\begin{figure}[t]
  \centering
  \includegraphics[angle=0, width=12cm]{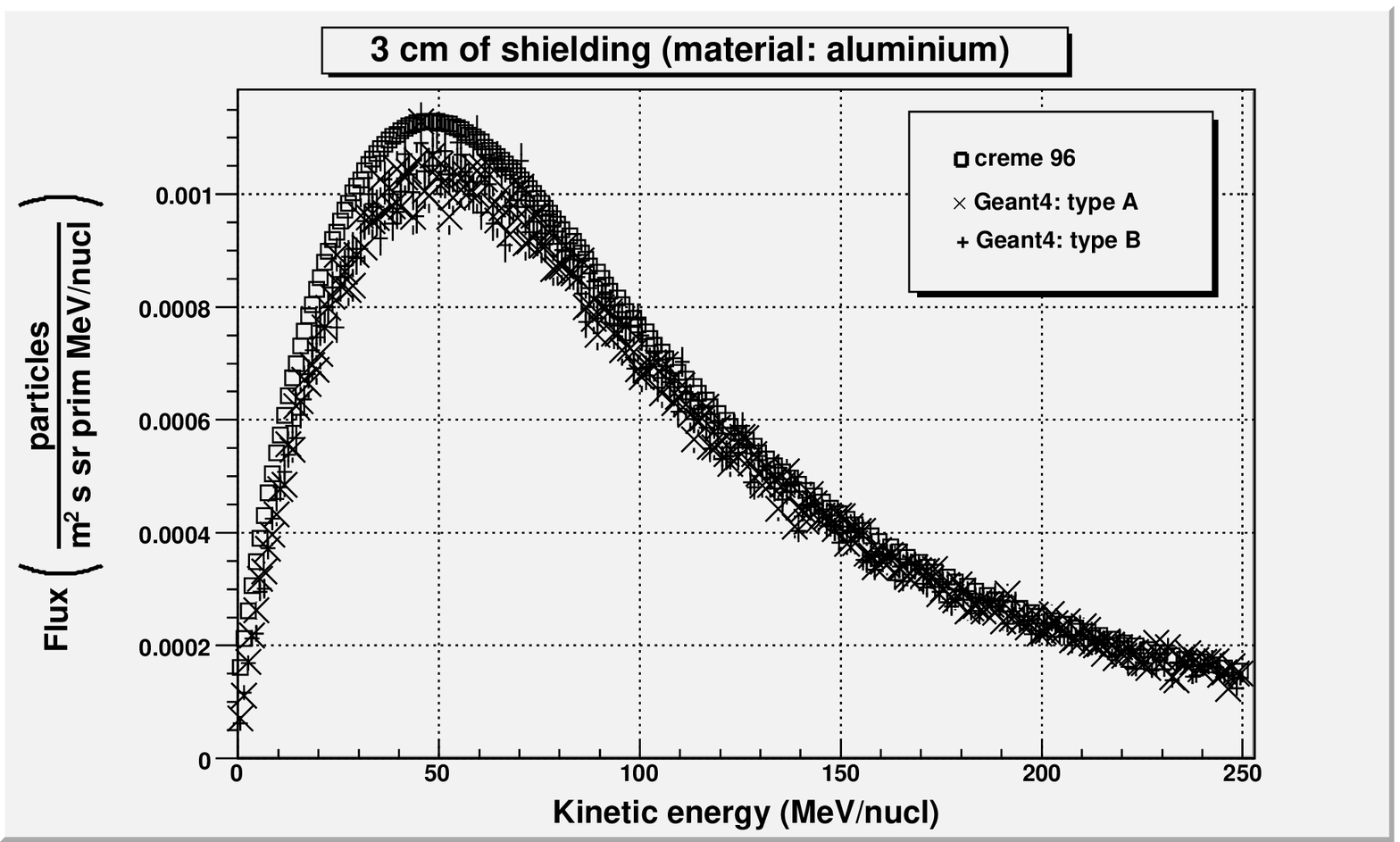}
  \caption{Proton fluxes behind 3~cm of aluminium shielding.
	   plotted in decimal scale.}\label{fig:3cm_dec}
\end{figure}

\begin{figure}[t]
  \centering
  \includegraphics[angle=0, width=12cm]{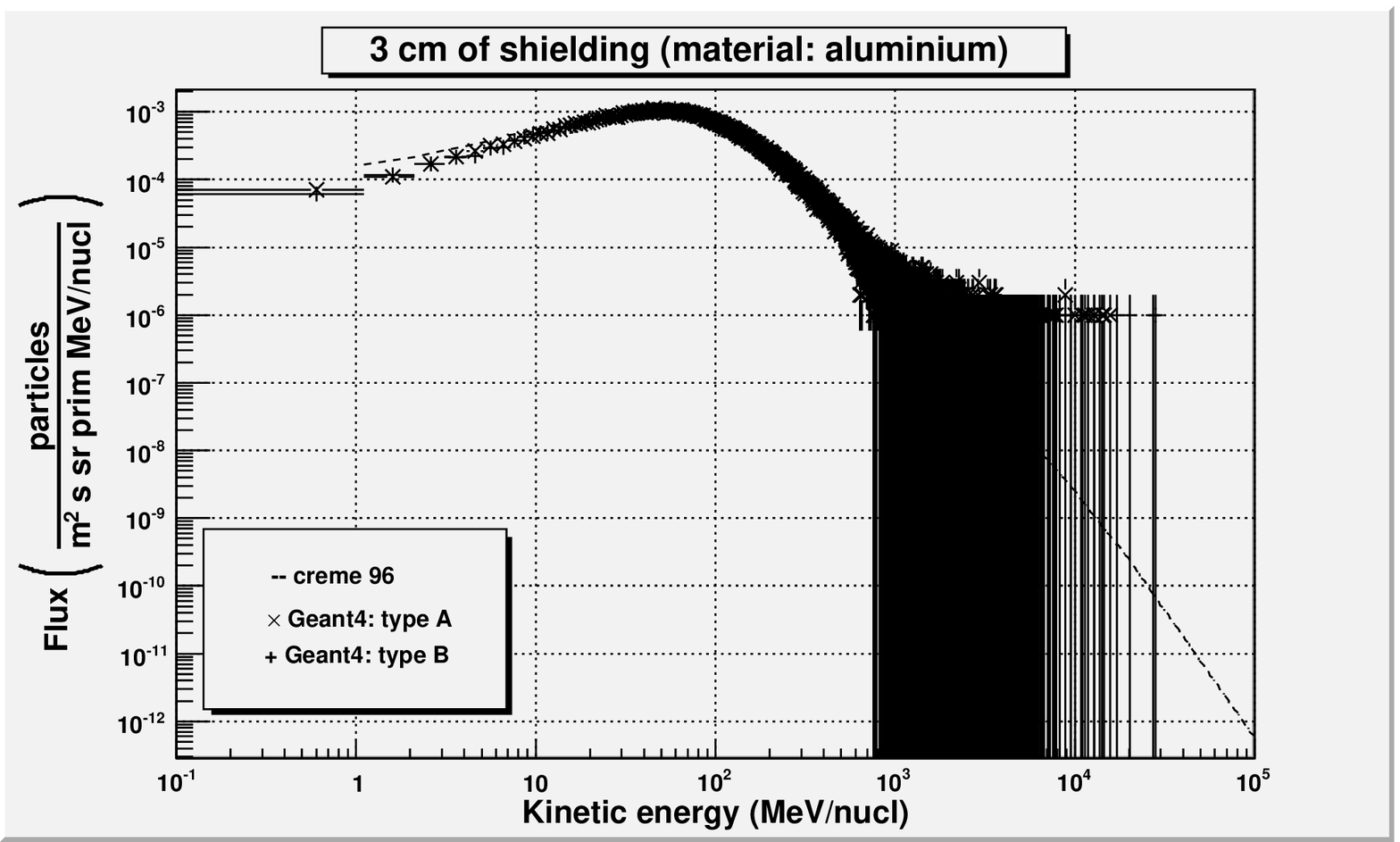}
  \caption{Proton fluxes behind 3~cm of aluminium shielding.
	   plotted in logarithmic scale.}\label{fig:3cm_log}
\end{figure}

\begin{figure}[t]
  \centering
  \includegraphics[angle=0, width=12cm]{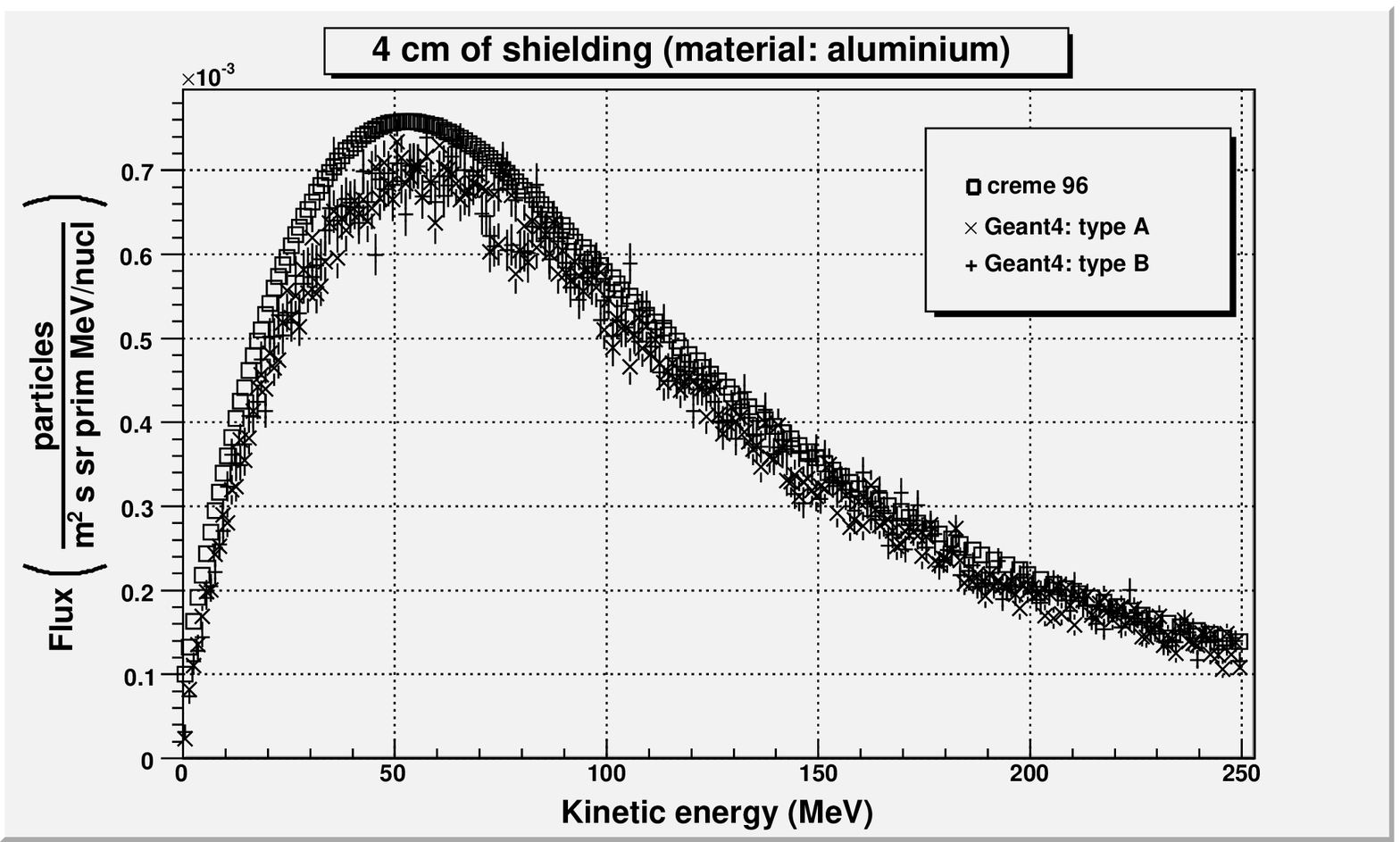}
  \caption{Proton fluxes behind 4~cm of aluminium shielding.
	   plotted in decimal scale.}\label{fig:4cm_dec}
\end{figure}

\begin{figure}[t]
  \centering
  \includegraphics[angle=0, width=12cm]{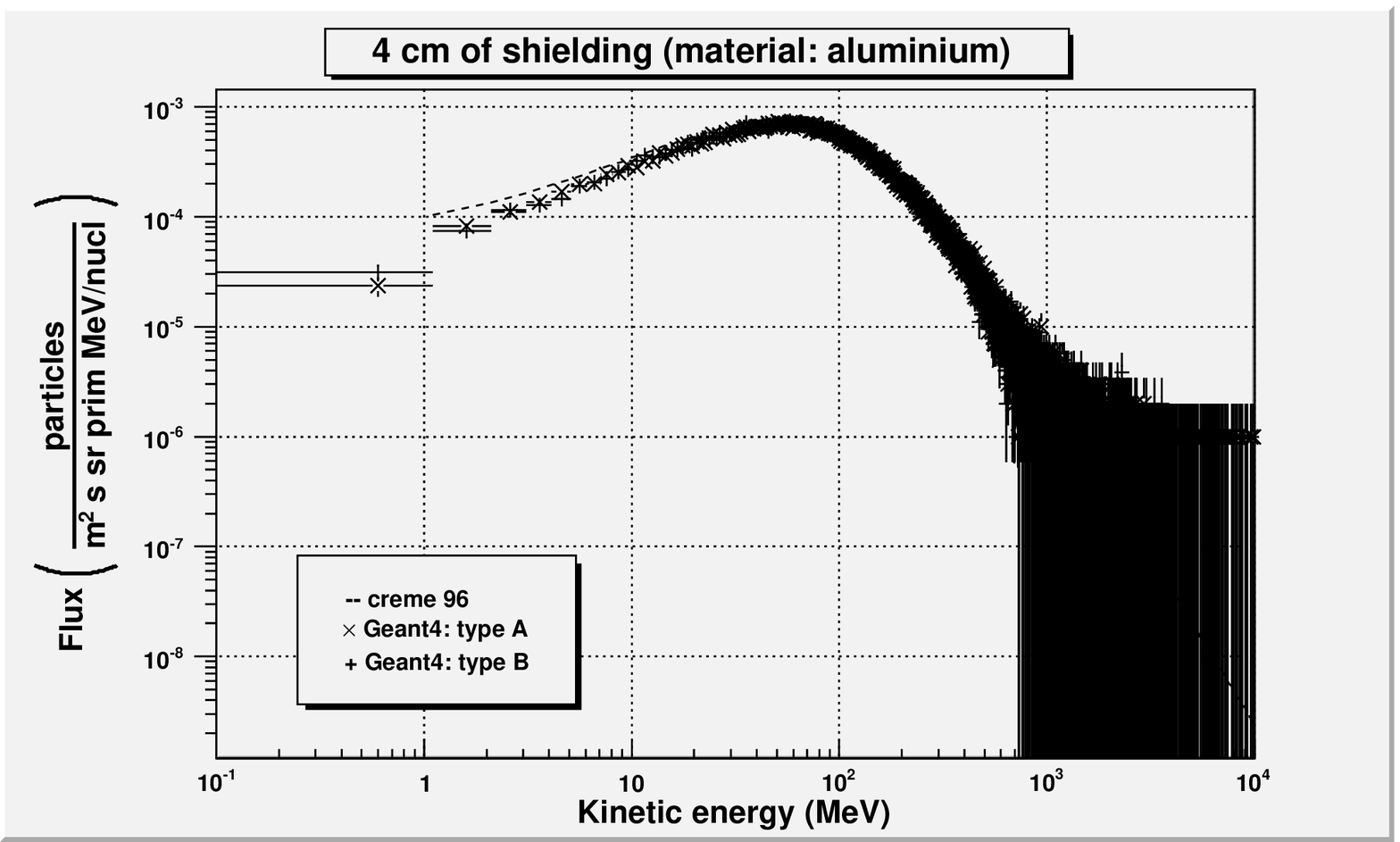}
  \caption{Proton fluxes behind 4~cm of aluminium shielding.
	   plotted in logarithmic scale.}\label{fig:4cm_log}
\end{figure}

\begin{figure}[t]
  \centering
  \includegraphics[angle=0, width=12cm]{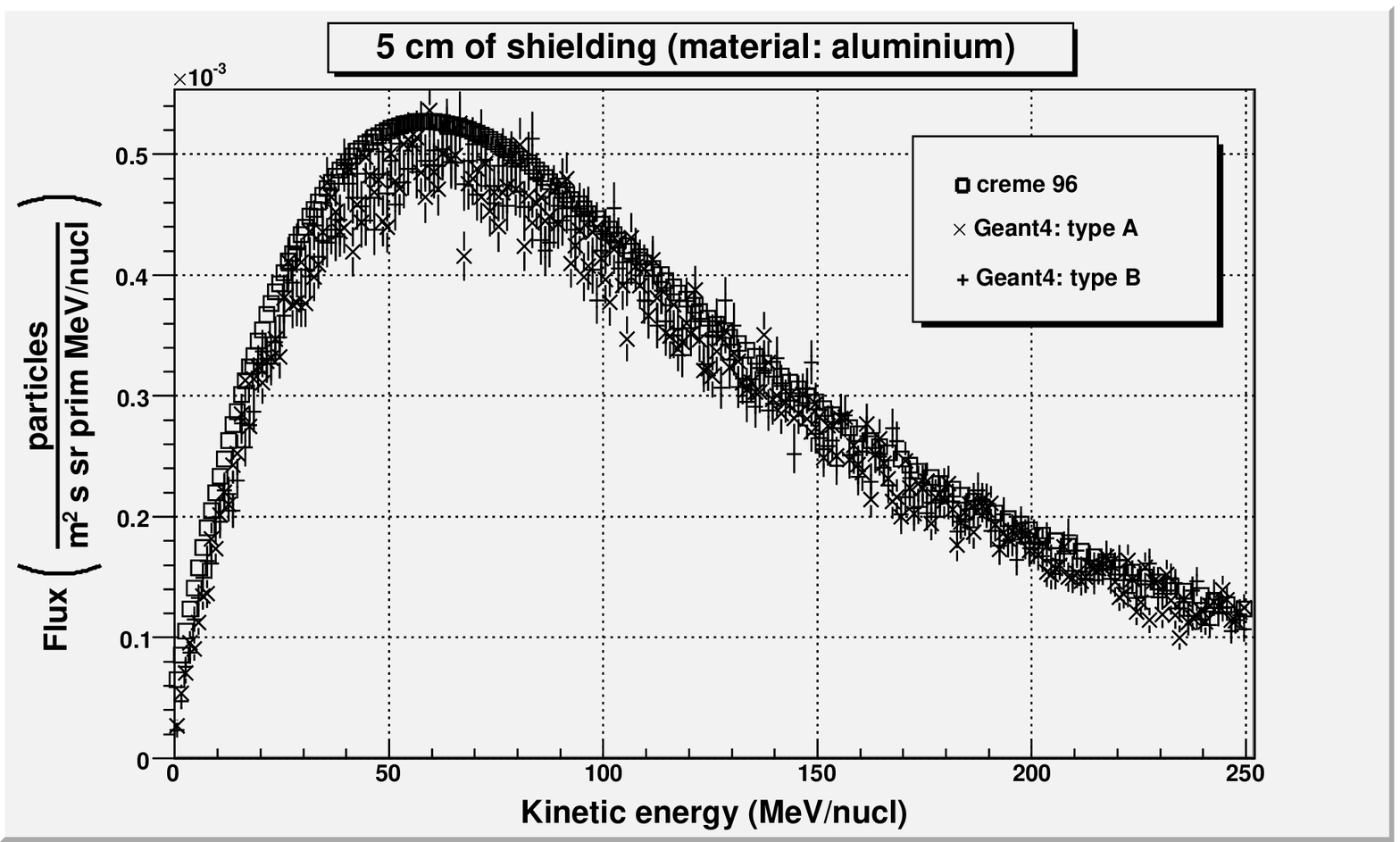}
  \caption{Proton fluxes behind 5~cm of aluminium shielding.
	   plotted in decimal scale.}\label{fig:5cm_dec}
\end{figure}

\begin{figure}[t]
  \centering
  \includegraphics[angle=0, width=12cm]{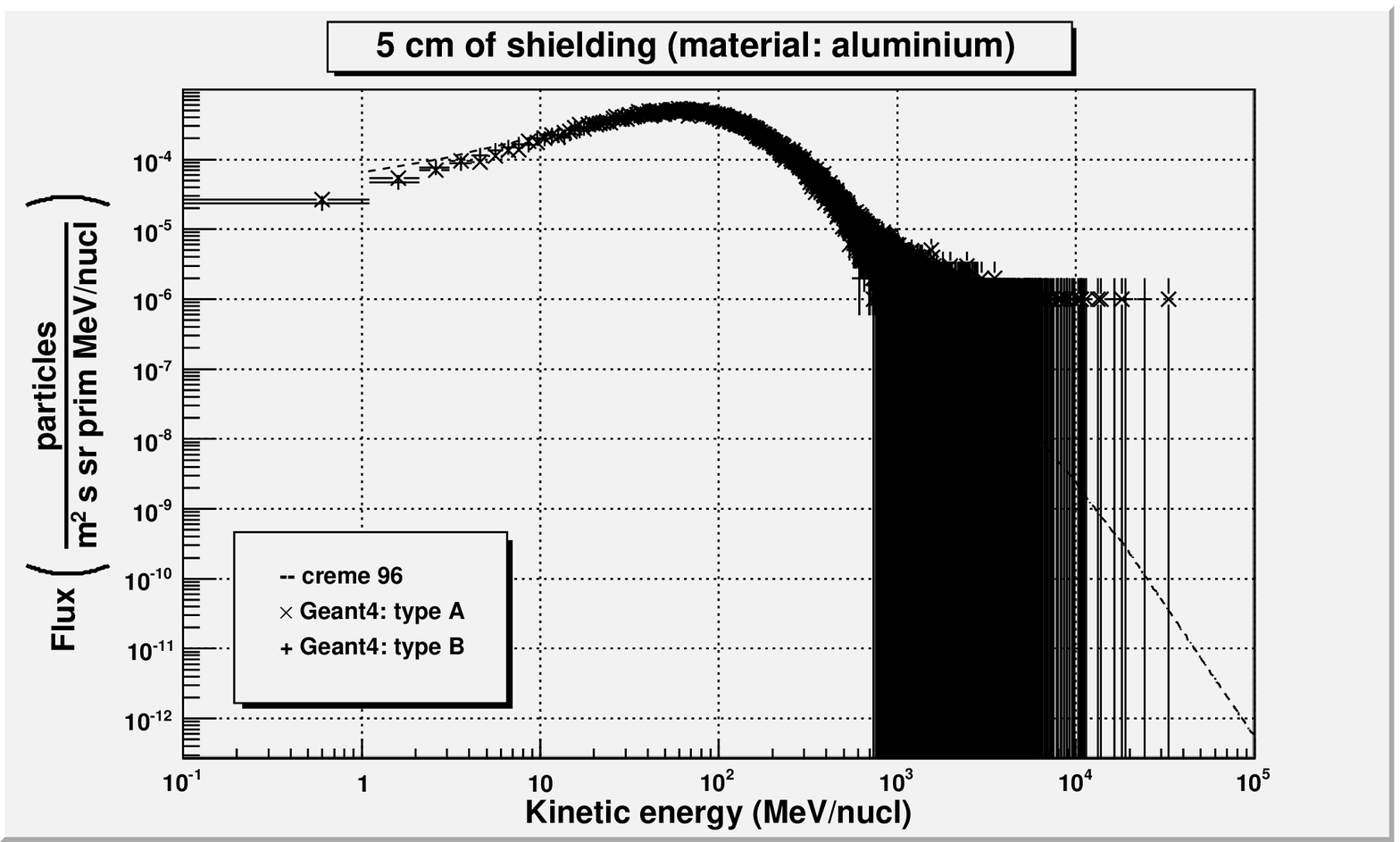}
  \caption{Proton fluxes behind 5~cm of aluminium shielding.
	   plotted in logarithmic scale.}\label{fig:5cm_log}
\end{figure}

Our results are plotted in fig.\ref{fig:1cm_dec}-\ref{fig:5cm_log} below  shielded proton fluxes are reported using two kinds of scale, decimal 
(figs.~\ref{fig:1cm_dec}, \ref{fig:2cm_dec}, \ref{fig:3cm_dec}, \ref{fig:4cm_dec}, \ref{fig:5cm_dec}) 
and logaritmic  
(figs.~\ref{fig:1cm_log}, \ref{fig:2cm_log}, \ref{fig:3cm_log}, \ref{fig:4cm_log}, \ref{fig:5cm_log}).

In the first case we can clearly observe the peak corresponding to particle energy with maximum transit probability.

%\noindent The aluminium shield blocks those protons with low kinetic energy, therefore by increasing the particle energy the flux drops rapidly.
%When primary particle flux goes nosedive, the rise of the shielded proton flux is stopped.
%These two phenomena produce the shape of the proton shielded flux.   

%\noindent By increasing the shielding thickness, peak levels change as particle energy with maximum transit probability (in the figures we cannot clearly see this value), because shielding cutting power is too strong.
%However, the number of peak particles are lower since the protons able to cross the shield are lower.
%Figs. \ref{fig:1cm_dec}, \ref{fig:2cm_dec}, \ref{fig:3cm_dec}, \ref{fig:4cm_dec}, \ref{fig:5cm_dec} show that the number of particles with maximum transit probability in terms of $10^{-4} \cdot \frac{particles}{m^2 \cdot s \cdot sr \cdot prim}$ decreases thus: 35, 18, 10, 7 and 5 as the thickness of the aluminium shield increases 1, 2, 3, 4 and 5~cm intervals. 

In the logarithmic version it is possible to view the energetic range of the particles and how the flux decreases. 
In this representation the error bars of the energetic particles are very wide because there are very low entries for the higher energies.
We maintain that the number of primary particles needed to obtain good error bars for high energy particles (w.r.t. our primary flux) should be more than $10^{10}$. 
This type of simulation is too expensive for our computational infrastructures, therefore we had to limit our simulation to $10^6$ primary particles. 

For all shielding thicknesses, from 1~cm to 5~cm, we can observe a good agreement between CREME96, and GEANT4 simulation data, for both physics list type. 
Sometime the overlapping of the different data is so good that we can not see the CREME96 data.
In figures where a decimal scale is used, we can distinguish the visible difference on the peak. 
This effect is due to different nature of the simulation code output data: CREME96 reports a continuous curve, GEANT4 shows the information data in a histogram therefore the flux shape is not a continuous curve.

\label{Results}

\section{Conclusion and future work}
We made a comparison between CREME96 and GEANT4, two particle transport codes through matter, in order to measure an aluminium target's outgoing proton fluxes and up until now evaluated using 
%%dealing with the protons fluxes outgoing from the aluminium target generated by the same primary proton flux, obtained using 
only CREME96.
To test GEANT4, we worked activating two different kinds of physics lists.
Our results show a good agreement between the output of the two simulators for each GEANT4 physics list considered.

These positive results encourage us to carry out more research in the same field in the future.
A possible evolution of this work is, in fact, the analysis using other primary particles, since CREME96 can consider particles with Z varying from 1 to 28. 
\label{FeaturesWorks}

\section{Acknowledgments}
We would like to thank all the people at Thales Alenia Space-Italy in particular
ing.~Cesare Lobascio, ing.~Mauro Briccarello, doc.~Roberto Destefanis and doc.~Emanuele Tracino.
This study was carried out during a scholarship activity program called Progetto Lagrange financed by the ISI and the CRT Foundation and Thales Alenia Space-Italy.   
\label{Acknowledge}

\bibliographystyle{unsrt}
\bibliography{biblio/biblio,biblio/creme,biblio/web}

\clearpage

\clearpage

\clearpage

\clearpage

\clearpage

\clearpage

\end{document}